\documentclass[prl,aps,amssym,nofootinbib,floatfix,twocolumn]{revtex4} 
\usepackage{graphicx,natbib}

\newcommand{\be}{\begin{equation}}
\newcommand{\ee}{\end{equation}}
\newcommand{\bea}{\begin{eqnarray}}
\newcommand{\eea}{\end{eqnarray}}
\newcommand{\pdag}{{\phantom{\dagger}}}
\begin{document}
\title{Non-Hermitian Luttinger Liquids and Vortex Physics}
\author{Walter Hofstetter$^1$, Ian Affleck$^2$, 
David R. Nelson$^1$, and Ulrich Schollw\"ock$^3$}
\affiliation{$^1$Lyman Laboratory, Harvard University, Cambridge, MA 02138, USA \\
$^2$Department of Physics, Boston University, Boston, MA 02215, USA \\
$^3$Sektion Physik, Universit\"at M\"unchen, Theresienstr. 37, 
D-80333 M\"unchen, Germany}
\date{\today}
\begin{abstract}
As a model of two thermally excited flux liquids connected by a weak
link, we study the effect of a single line defect on vortex
filaments oriented parallel to the surface of a thin planar
superconductor.  
When the applied field is \emph{tilted} relative to the line defect,  
the physics is described by a \emph{nonhermitian} 
Luttinger liquid of interacting quantum bosons in one spatial 
dimension with a point defect.   We analyze this
problem using a combination of analytic and numerical density
matrix renormalization group methods, uncovering a delicate
interplay between enhancement of pinning due to Luttinger liquid
effects and depinning due to the tilted magnetic field.
Interactions dramatically improve the ability of a single columnar pin 
to suppress vortex tilt when the Luttinger liquid parameter $g\le1.$
\end{abstract}
\maketitle

\section{Introduction}
The past decade has seen considerable work on the statistical 
mechanics and dynamics of thermally excited vortices in type II 
high--temperature superconductors \cite{Blatter94}.
The competition between interactions, pinning and thermal fluctuations 
gives rise to a wide range of novel phenomena, including a first order 
melting transition of the Abrikosov flux lattice into an entangled 
liquid of vortex filaments \cite{Nelson88} 
and a low--temperature \emph{Bose glass} phase with vortices strongly pinned 
to columnar defects \cite{Nelson92}.

A convenient way of understanding interacting flux lines is provided by 
the formal mapping between the classical statistical mechanics of (d+1)--dimensional 
directed flux lines and the nonrelativistic quantum mechanics of d--dimensional 
bosons. 
In this mapping, flux lines traversing the sample along the 
direction of the external magnetic field $\mathbf{H} = H \hat{\bf z}$  
correspond to boson world lines propagating in imaginary time $\tau$. 
The classical partition function of thermally excited vortex lines 
is proportional to a quantum mechanical matrix element. 
The thickness of the sample in the $z$--direction, $L_z$, corresponds to  
the inverse temperature $\beta \hbar$ of the bosons, while 
thermal fluctuations of the vortices, due to finite $k_B T$, play the
role of quantum fluctuations of the bosons, controlled by $\hbar$.

If the direction of the external magnetic field does \emph{not} coincide 
with $\hat{\bf z}$, the direction of 
the columnar defects, it is convenient to separate the transverse 
component of the field $H_\perp$ from the parallel one $H_{\parallel}$ along 
$\hat{\bf z}$. When $H_\perp \ll H_{\parallel}$, the transverse component 
$H_\perp$ plays the role of a constant imaginary vector potential 
for the bosons \cite{Nelson92,Hatano}. The corresponding quantum Hamiltonian is 
\emph{non--hermitian}, with new and interesting properties. 
Stimulated by vortex physics, there has been considerable interest in 
models of noninteracting quantum particles in a constant imaginary vector potential 
with a disordered pinning potential. 

Less is known about non--hermitian models \emph{with} interactions. 
A disordered array of parallel columnar defects leads to a strongly pinned 
low--temperature Bose glass phase. For $H_\perp$ less than a critical value $H^c_\perp$,  
this phase exhibits a ``transverse Meissner effect'', such that the 
vortex filaments remain pinned parallel to the columns even though the external field 
is tilted away from the column direction. 

\begin{figure}[h]
\includegraphics[width=0.7\linewidth]{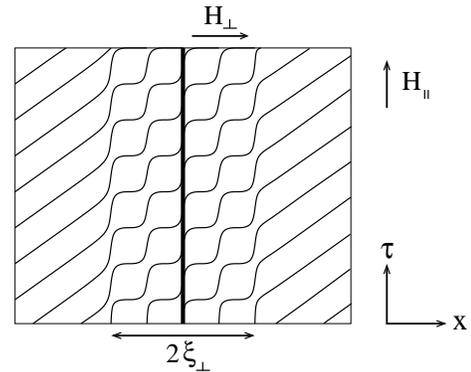}
        \caption{\label{fig:setup}
Schematic snapshot of vortex lines and a columnar pin. The near--vertical segments occur at 
locations of maxima of the average density. For finite tilt these maxima 
become negligible for $|x|>\xi_\perp$. 
Here, the $z$--axis is denoted by $\tau$.}
\end{figure}

In this paper we study the effect of a \emph{single} columnar pin 
(or an equivalent linear defect) on the 
statistical mechanics of thermally fluctuating vortex lines confined 
in a thin, superconducting slab (see Fig. \ref{fig:setup}). 
With exception of a brief discussion in \cite{Nelson92} 
and the analysis of Ref.~\cite{Radzihovsky}, little has been done on vortex 
physics in the limit of a dilute concentration of columns (or twin planes). 
If the average spacing between defects is $d$, this is the regime 
$H_\parallel \gg B_\Phi = \Phi_0/d^2$ where $B_\Phi$ is the ``matching field'' 
and $\Phi_0$ is the flux quantum. 
The large number of ``interstitial'' vortices 
between defects could be locally crystalline or melted into a flux liquid. 
A third possibility (perhaps the situation most closely related to our 
calculations in 1+1 dimensions) 
is vortices in a ``supersolid'' phase, with extended correlations 
in both the translational and boson order parameters \cite{Frey94}.

The feasibility of studying vortex physics in samples which are effectively 
(1+1)--dimensional was demonstrated by Boll\'e et al. \cite{Bolle} in thin samples 
of NbSe$_2$, where the effect of point disorder on interacting vortices 
near $H_{c1}$ was observed. 
Similar experiments might be possible on thin high--$T_c$ samples where 
columnar defects could be implemented mechanically by cutting a thin ``notch''. 
A related problem in (2+1) dimensions concerns the effect of a single twin plane 
or grain boundary on vortex matter, where point disorder may lead to 
algebraic decay of density correlations \cite{Giamarchi} of the Abrikosov flux 
lattice similar to those discussed below. A single such plane has 
a similar pinning effect on bulk flux lines as does a linear defect 
in (1+1)--dimensional systems.



In the following we will consider a single columnar defect in a system 
of interacting flux lines in (1+1) dimensions. 
Although  results can also be obtained using a flux--line related 
phonon formalism \cite{Hwa93}, we here found it convenient 
to work with an equivalent quantum Hamiltonian \cite{Blatter94,Hatano}
\bea  \label{Hamcon}
\hat{H} &=& - \frac{(k_B T)^2}{2m} \int dx \Psi^\dagger(x) 
\left(\frac{d}{dx} - h \right)^2 \Psi(x)   \\
&& + \frac{1}{2} \int dx dy \, n(x) V\left(|x-y|\right) n(y) - \epsilon_0 \, n(0) \nonumber
\eea
where $V(|x|)$ is a short--range repulsive vortex interaction potential, 
$\Psi(x)$ annihilates a bosonic flux line, $n(x)= \Psi^\dagger(x) \Psi(x)$
is the boson number density and $m$ is the vortex tilt modulus.
The imaginary vector potential  
$h= \Phi_0 H_\perp / (4 \pi T)$ arises due to the tilted magnetic field 
and $\epsilon_0$ is the strength of the defect modelled by a $\delta$--potential 
at the origin. In the following we set $k_B T=1$ 
(i.e. $\hbar=1$ in the quantum model).  

Without the local potential and the non--hermitian term this model has been 
well studied \cite{Lieb,Haldane}. In particular, Haldane \cite{Haldane} 
has shown that this spinless Luttinger liquid exhibits a line of critical points with 
continuously varying exponents. His calculation is based on the bosonization technique,  
where the boson field 
\be
\Psi^\dagger \sim \sqrt{n_0 +\frac{du}{dx}} 
\sum_{m=-\infty}^{\infty} e^{i 2\pi m (n_0 x + u(x))} e^{i\phi(x)}
\ee 
is represented in terms of a boson phase operator $\phi(x)$ and (dimensionless) 
phonon operator $u(x)$. The two fields satisfy the commutation relation
$ [\phi (x), u(y)]=( i/2) \ \hbox{sgn} (x-y)$.
With applications to vortex physics in mind, 
we have extended the bosonization approach 
and work on quantum impurities \cite{Kane,Eggert} to the non--hermitian case $h>0$ 
and calculated asymptotic low--energy properties for the model (\ref{Hamcon}).

In addition, we have performed a non--perturbative numerical analysis 
using the Density--Matrix Renormalization Group (DMRG) \cite{White} 
for a discretized version of the Hamiltonian (\ref{Hamcon})  
\bea \label{Hamdisc}
H &=& \sum_{i=0}^L \big[-t \left(b^\dagger_i b^\pdag_{i+1} e^h + 
b^\dagger_{i+1} b^\pdag_i e^{-h}\right) + \frac{U}{2} n_i (n_{i}-1) \nonumber \\
&&+ V n_i n_{i+1}\big] - \epsilon_0 b^\dagger_0 b^\pdag_0
\eea
corresponding to a nonhermitean \emph{Bose--Hubbard} model where 
$n_i=b^\dagger_i b^\pdag_i$ and the hopping is 
$t=1/2m$ (for unit lattice constant). In the following we set $t=1$. 
We work in the canonical ensemble, fixing the density of bosons per site $n_0$.  
We have retained an onsite and a next--neighbor interaction, which turn out to be 
sufficient to qualitatively describe the full phase diagram. 
Furthermore, for computational purposes we allow at most 2 bosons per site, 
which effectively renormalizes the on--site repulsion. 
The lattice model (\ref{Hamdisc}) is a good approximation to (\ref{Hamcon}) for 
small filling $n_0$ (average number of bosons per site). 
Our calculation is based on an extension of the DMRG to non--hermitian 
systems with complex eigenvalues and eigenvectors 
(for details see \cite{Carlon}).

\begin{figure}[h]
\includegraphics[width=0.76\linewidth]{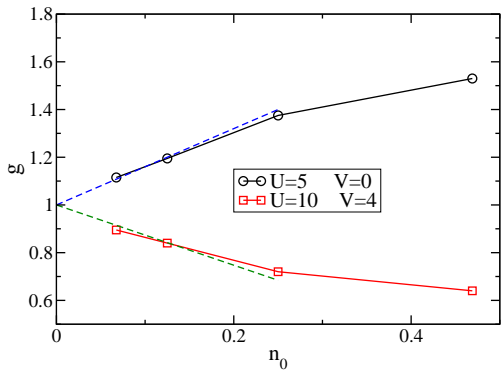}
        \caption{\label{fig:eta_vs_density}
Luttinger--liquid parameter $g$ vs. density. 
The dashed lines show the analytic result 
at low densities.}
\end{figure}

In the hermitian case $h=0$ without impurity we have first calculated  
the Luttinger liquid parameter $g$ which governs the long--wavelength 
behavior of correlation functions and is important to understand 
the response at finite tilt. 
We adapt the DMRG work of Ref.~\cite{Kuehner} to  
\emph{periodic} boundary conditions, essential for the 
study of persistent currents (i.e.~arrays of tilted vortex lines)
discussed below. 
We focus exclusively on the superfluid (Luttinger liquid) phase.  
Via DMRG we have calculated the 
boson correlation function, which from conformal field theory 
is expected to behave as 
$ <\Psi^\dagger (x)\Psi (0)> \sim  |L \sin(\pi x/L)|^{-1/2g}$.
We have verified this behavior numerically with high accuracy 
and have extracted $g$ by a fit to the data 
(see Fig.~\ref{fig:eta_vs_density}). 
We also determined $g$ from the compressibility 
and the finite-size dependence of the ground state energy \cite{c}  
with excellent agreement between the values of $g$ obtained 
by both methods. 
For arbitrary short range potentials (in continuum or lattice models) we have derived the
general low--density result $g\approx 1-2an_0+O(a^2n_0^2)$ where $a$ is the
two-particle scattering length. For our lattice Hamiltonian (\ref{Hamdisc}) we find 
$a=-(8t^2-4tV-UV)/(2tU+UV+4tV)$.
As shown in Fig.~\ref{fig:eta_vs_density}, 
this asymptotic result is in good agreement with the numerical data. 

We now include the pinning term proportional to $\epsilon_0$ in 
Eq.~(\ref{Hamcon}). 
In order to determine the relevance of this term at long wavelengths 
we have performed a perturbative renormalization group (RG) analysis.  
We obtain the following renormalization flow of the pinning strength 
\be
\epsilon_0(l)=\epsilon_0(l_0)\left({l_0\over l}\right)^{g-1} \label{epeff}
\ee
where $l$ is an effective length scale or inverse cutoff momentum. 
For $g>1$ the renormalized coupling to flows to zero at 
long length scales while for $g<1$ it diverges. 

Remarkably, while in fermionic systems with (generic) repulsive interactions 
one always has $g<1$, the \emph{bosonic} Luttinger liquid studied here 
can be tuned to either regime.
This can be easily seen by setting $U=\infty, V=0$ in (\ref{Hamdisc}): 
Since hardcore bosons in 1d are equivalent to 
noninteracting spinless fermions, we obtain $g=1$. 
Smaller $U$ increases $g$ from 1 while additional next--neighbor interactions 
$V>0$ lead to $g<1$.
In the following we will denote the special situation $g=1$ 
as the \emph{free fermion limit}, for which we have replaced 
the DMRG by computationally less expensive exact diagonalization.

The irrrelevance/relevance of the pin can be clearly observed 
in the Friedel oscillations of the boson density 
$\Delta n(x) \equiv \langle n(x)\rangle - n_0$, for which we find the 
analytic result 
\be \label{Friedel2}
\Delta n(x) \sim \frac{\cos(2\pi n_0 |x|)}{|x|^\alpha} \; e^{-|x|/\xi_\perp(h)}.
\ee
with an interaction--dependent exponent 
$\alpha= \left\{g \atop 2g-1\right\}$ for $\left\{g<1 \atop g>1\right\}$
and a \emph{decay length} $\xi_\perp \sim 1/h$.
At zero tilt $h$ we therefore find a pure power--law decay of 
the density perturbation which is more rapid 
when the pin is irrelevant.
When $h>0$, the Friedel oscillations acquire an additional 
exponential decay. 

\begin{figure}[h]
\includegraphics[width=0.8\linewidth]{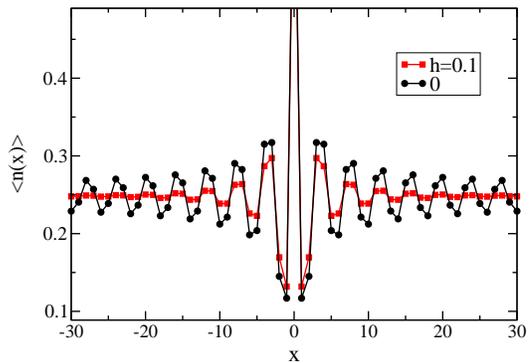}
        \caption{\label{fig:friedel}
Friedel oscillations of the flux line density vs. distance $x$ from the defect, 
calculated numerically for $U=10$, $V=4$, $L=128$ and $n_0=0.25$, corresponding to $g\approx 0.72$.}
\end{figure}
As illustrated in Fig.~\ref{fig:friedel}, the density oscillates 
with a phase set by the impurity position, and an algebraic envelope 
before exponential decay sets in for $x>\xi_\perp(h)$. 
In the vortex picture (see Fig.~\ref{fig:setup}), configurations are dominated 
by parallel, tilted flux lines at distances larger than $\xi_\perp(h)$ 
from the pin. Closer in, vortices attempt to align with the maxima 
in the density oscillations present when $h=0$. This alignment 
is limited by interactions as vortices enter and leave the aligned region 
with increasing imaginary time $\tau$. 
The resulting vortex configurations resemble a symmetric traffic jam,  
with vortices queuing up (and occasionally changing places) 
in the vicinity of the columnar defect. With our conventions, the slope 
of the lines far from the pin is $h/m$, so new vortices enter the jam 
at imaginary time intervals $\xi_\parallel \approx m/h n_0$, where $n_0$ 
is the linear density of ``bosons''. 
If $c$ is the Luttinger liquid velocity 
we expect that $\xi_\perp(h) \sim c\, \xi_\parallel \sim 1/g h$, a diverging 
length scale we confirm with our analytic calculations. 
Note that the pinning strength is reduced dramatically for length scales 
$x>\xi_\perp(h)$ even for $g<1$.

When $h>0$, the non--hermiticity leads to  
a finite persistent current $J_b = -\frac{i}{L} d\langle \hat{H} \rangle/dh$  
in the ground state. 
This current is purely imaginary and corresponds to the 
transverse magnetization in the original flux line system \cite{Hatano}: 
\be
M_\perp \sim \Phi_0 {\rm Im}J_b.
\ee
The defect reduces this current, 
due to flux lines pinned even in the 
presence of a tilted magnetic field. 
Although a single pin cannot modify the bulk current 
in the thermodynamic limit $L\to\infty$, 
it creates nontrivial finite--size effects. 
Since ${\rm Im} J_b=h N_b/m L$ in the absence of pinning (where $N_b\equiv n_0 L$), 
it is convenient to define a ``pinning number'' $N_p$ for vortices given by
${\rm Im} J_b \equiv h (N_b - N_p)/mL$. 
Because ${\rm Re} J_b=0$, this can be written   
\be
N_p \equiv N_b 
[J_b(0)-J_b(\epsilon_0 )]/J_b(0)
\ee

\begin{figure}[h]
\includegraphics[width=0.8\linewidth]{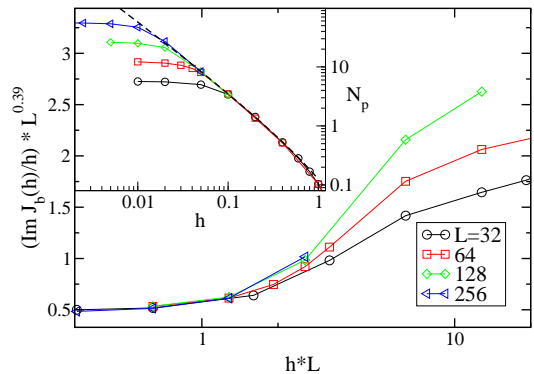}
        \caption{\label{fig:linear_current}
Main plot: Finite--size scaling of the current (DMRG results) for filling $n_0=0.25$ 
and a relevant pin ($g\approx 0.72$). 
Notice the data collapse in the linear--response regime $hL\to 0$.
Inset: pinning number $N_p$ as a function of $h$ for the same parameters. 
The dashed line gives the logarithmic behavior in Eq. (\protect{\ref{log}}) 
with an offset of $const. = 0.5$.
}
\end{figure}

The quantity $N_p$ may be readily calculated for the free fermion case $g=1$ 
where the groundstate energy is determined by filling up all the states 
below the ``Fermi surface''. 
The pinning number obtained in this way has the asymptotic behaviour $(L\to \infty)$  
\bea N_p &\to& 
{(m\epsilon_0)^2\over 2\pi^2n_0h},\ \  (m\epsilon_0<<h) \nonumber \\
&\to & {n_0\over h}\ln (|\epsilon_0 |/n_0),\ \  (m\epsilon_0>>h), \label{asymptote}
\eea
results valid provided $h \ll n_0 = N_b /L$.
Remarkably, $N_p$ diverges as $h\to 0$. When the pinning is strong, 
the functional form (\ref{asymptote}) can be understood in terms of 
the aligned local density wave which extends out to a distance $\xi_\perp$.  
The $N_p \approx \xi_\perp n_0 \approx n_0/h$ vortices entrained 
in this ``traffic jam'' do not contribute to the current. 
\begin{figure}[h]
\includegraphics[width=0.78\linewidth]{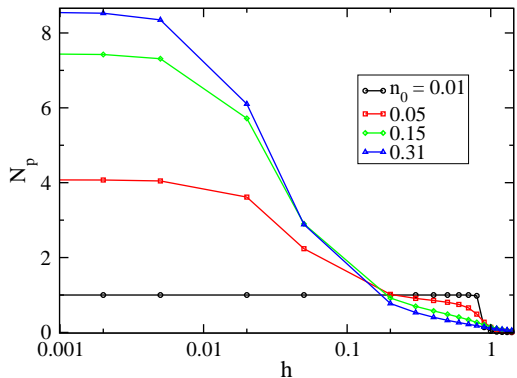}
        \caption{\label{fig:pinning_fraction1}
Pinning number in the free fermion limit $g=1$ 
for $L=100$ and $\epsilon_0=2$. 
Note the ``step'' due to single--vortex depinning and the 
strong enhacement at small tilt $h$.
}
\end{figure}
To check this divergence, we have also calculated the pinning number  
numerically within the lattice model (\ref{Hamdisc}), both in the free fermion limit 
($U=\infty, V=0$) and for general interactions. 
Results are shown in Fig.~\ref{fig:pinning_fraction1}.
A clearly visible feature is the ``step'' at intermediate tilt 
for low boson densities, corresponding to the single--vortex 
depinning transition at $h_c \approx m \epsilon_0$ \cite{Hatano}. 
Most prominent, however, is the dramatic increase in the number 
of pinned vortices at small tilt. 
We find similar results with DMRG for $g<1$.
In the \emph{linear response} limit ($hL\to 0$) we find more generally that 
$N_p \sim L^{3-2 g}$ for an irrelevant defect ($g>1$), while 
in the relevant case almost all vortices are pinned, 
i.e. $N_p \to N_b$ for large system size $L$. 
The \emph{residual} current for a relevant pin has the linear response 
form $J_b(h)|_{h\to 0} \sim h L^{1 - 1/g }$, which vanishes as $L\to \infty$. 
A similar result for fermions was obtained by Gogolin and Prokov'ev \cite{Gogolin}
in the case of a real vector potential. 
We have verified this finite--size dependence of the current 
with good accuracy in the DMRG calculation (see Fig.~\ref{fig:linear_current}).

The equivalence to a real vector potential 
breaks down at finite $h$. Our results are consistent with 
two qualitatively different 
types of behaviour depending on the value of $g$. While for an irrelevant pin 
($g>1$) a simple power--law scaling ansatz 
$ N_p(h) = h^{-3+2g} \Phi(h L)$
works, our analytic work suggests 
a nontrivial logarithmic correction for $g<1$:
\be \label{log}
N_p(h) = (n_0/h) \left(-\left(1/g - 1\right) \ln(h) 
+ {\rm const.}\right)
\ee
This equation is valid for $hL \gg 1$ and $h \ll n_0$.
The DMRG data (see inset of Fig.~\ref{fig:linear_current}) 
are consistent with this conjecture.
%
%

In conclusion, we have studied the effect of a single columnar defect 
on a sea of interacting vortices in 1+1 dimensions, 
in the presence of a tilted magnetic field. 
The physics is described in terms of the groundstate 
of a nonhermitian Luttinger liquid.  
Our calculations demonstrate that repulsive interactions can lead to a 
dramatic enhancement in the number of pinned flux lines for $g<1$ 
and thus to a strong transverse Meissner effect 
controlled by $\xi_\perp(h)$.  
Details of our analytic and numerical work will appear later 
\cite{long_paper}.

Acknowledgements: We would like to acknowledge discussions on the
experimental situation with M. Marchevsky and E. Zeldov, and  
conversations with L. Radzihovsky. 
Work by W.H. and D.R.N. was supported by the National Science Foundation through
Grant DMR-0231631 and the Harvard Materials Research Laboratory via Grant
DMR-0213805.  W.H. and U.S. also acknowledge financial support 
from the German Science Foundation (DFG). 
Work by I.A. was supported by the National Science Foundation 
through Grant DMR-0203159.

\pagebreak







\begin{thebibliography}{999}
\bibitem{Blatter94}G. Blatter \emph{et al.}, Rev. Mod. Phys. \textbf{66}, 1125 (1994).
\bibitem{Nelson88}D.R. Nelson and H.S. Seung, Phys.  Rev.  {\bf B39}, 9153 (1989).
\bibitem{Nelson92}D.R. Nelson and V.M. Vinokur, Phys. Rev. {\bf B48}, 13060 (1993).
\bibitem{Hatano}N. Hatano and D.R. Nelson, Phys. Rev. {\bf B56}, 8651 (1997).
\bibitem{Radzihovsky}L. Radzihovsky, Phys. Rev. Lett. {\bf 74}, 4923 (1995). 
\bibitem{Frey94}E. Frey \emph{et al.}, Phys. Rev. {\bf B49}, 9723 (1994).
\bibitem{Bolle}C.A. Boll\'e \emph{et al.}, Nature {\bf 399}, 43 (1999).
\bibitem{Giamarchi}T. Giamarchi and P. Le Doussal, Phys. Rev. Lett. 
{\bf 72}, 1530 (1994). 
\bibitem{Hwa93}T. Hwa \emph{et al.}, Phys. Rev. {\bf B48}, 1167 (1993).
\bibitem{Lieb} E.H. Lieb and W. Liniger, Phys. Rev. {\bf 130}, 1605 (1963). 
\bibitem{Haldane}F.D.M. Haldane, Phys. Rev. Lett. {\bf 47}, 1840 (1981).
\bibitem{Kane} C.L. Kane and M.P.A. Fisher, Phys. Rev. {\bf B46}, 15233 (1992).
\bibitem{Eggert} S. Eggert and I. Affleck, Phys. Rev. {\bf B46}, 10866 (1992).
\bibitem{White}S.R. White, Phys. Rev. Lett. {\bf 69}, 2863 (1992).
\bibitem{Carlon}E. Carlon \emph{et al.}, Eur. Phys. J. {\bf B12}, 99 (1999).
\bibitem{Kuehner}T. K\"uhner \emph{et al.}, Phys. Rev. {\bf B61}, 12474 (2000).
\bibitem{c} H.W.J. Bl\"ote \emph{et al.}, Phys. Rev. Lett. {\bf 56}, 742 (1986); 
I.~Affleck, Phys. Rev. {\bf 56}, 746 (1986).
\bibitem{Gogolin} A. Gogolin and N. Prokof'ev, Phys. Rev. {\bf B50}, 4921
 (1994).
\bibitem{long_paper}I. Affleck, W. Hofstetter, D.R. Nelson, and U.~Schollw\"ock, 
to appear. 


\end{thebibliography}
\end{document}